\documentclass[prb,showpacs,byrevtex,floatfix,twocolumn]{revtex4}
\usepackage{amssymb,graphicx,afterpage}
\begin{document}
\title{Magneto-optics induced by the spin chirality in itinerant ferromagnet
Nd$_2$Mo$_2$O$_7$}
%
%
\author{I. K\'{e}zsm\'{a}rki$^1$, S. Onoda$^2$, Y. Taguchi$^3$, T. Ogasawara$^4$, M. Matsubara$^1$, S. Iguchi$^1$, N. Hanasaki$^1$,  N. Nagaosa$^{1,4}$, and Y. Tokura$^{1,2,4}$}

\affiliation{
$^1$Department of Applied Physics, University of Tokyo, Tokyo 113-8656, Japan}
\affiliation{
$^2$Spin Superstructure Project, ERATO, Japan Science and Technology Agency,
c/o Department of Applied Physics, University of Tokyo, Tokyo 113-8656, Japan}
\affiliation{
$^3$Institute for Materials Research, Tohoku University, Sendai 980-8577, Japan}
\affiliation{
$^4$Correlated Electron Research Center (CERC), National Institute of Advanced Industrial Science and Technology (AIST), Tsukuba 305-8562}
\date{version \today}
%
%
\pacs{78.20.Ls,
78.20.Bh,
}
\begin{abstract}
It is demonstrated both theoretically and experimentally that the
spin chirality associated with a noncoplanar spin configuration
produces a magneto-optical effect. Numerical study of the two-band
Hubbard model on a triangle cluster shows that the optical Hall conductivity
$\sigma_{xy}(\omega)$ is proportional to the spin chirality.
The detailed comparative experiments on pyrochlore-type molybdates
$R_2$Mo$_2$O$_7$ with $R=$Nd (Ising-like moments)
and $R=$Gd (Heisenberg-like ones) clearly distinguishes the two mechanisms,
i.e., spin chirality and spin-orbit interactions.
It is concluded that for $R$=Nd, $\sigma_{xy}(\omega)$ is
dominated by the spin chirality for the dc ($\omega=0$) and the $d \to d$
incoherent intraband optical transitions between Mo atoms.
\end{abstract}
\maketitle

%
%


Quantum-mechanical phenomena associated with the Berry
phase~\cite{Berry} have attracted much interest in
condensed-matter physics, including charge and spin Hall transport
properties. In particular, it has recently been recognized that
the Berry phase is relevant to the anomalous Hall effect (AHE)
where a finite dc Hall conductivity
$\sigma_{\rm Hall}(\equiv\sigma_{xy}(\omega=0))$ appears in ferromagnets.
It was first proposed that the AHE originates from a fictitious magnetic
flux penetrating the conducting network~\cite{Ong98,Ye99,Lyanda-Geller01}.
However, the AHE is distinct from the ordinary Hall effect that
appears in the presence of the external uniform magnetic field,
since the periodicity of the crystal is preserved in the case of the AHE.
Then it was found that the Berry phase curvature of the Bloch wavefunction
in the momentum space is a more fundamental and universal
concept~\cite{Ohgushi00,Taguchi01,MOnoda02,Fang03}, which leads to
a finite anomalous contribution to the Hall conductivity.
This Berry phase curvature emerges through
the relativistic spin-orbit interaction and/or the solid angle
subtended by the spins under the noncoplanar structure, the latter
of which is called {\it spin chirality}~\cite{BaskaranAnderson88,
Laughlin88,WenWilczekZee89,LeeNagaosa92}.

The spin-chirality mechanism for the AHE has a fundamental significance,
since the spin-orbit coupling acting on the propagating carriers was
considered to be indispensable to produce the AHE in conventional
theories~\cite{KarplusLuttinger54,Kondo62}.
Recently, spin-chirality scenarios for the AHE have been extensively
studied from both theoretical and experimental
aspects~\cite{Ong98,Ye99,Lyanda-Geller01,Ohgushi00,Taguchi01,SOnoda03,Taguchi03,Iguchi04,Tatara02}.
On the other hand, it is also believed that the spin-orbit coupling is
indispensable to have a finite optical Hall conductivity
$\sigma_{xy}(\omega)$.
It remains as an open issue how such magneto-optical properties are
influenced by the spin chirality in strongly correlated electron
systems like transition-metal oxides,
which is of our interest in the present paper.

In conventional magnets, it is difficult to distinguish between
two mechanisms based on the spin-orbit interaction and the spin
chirality, since the noncoplanar spin structure itself originates
mostly from the spin-orbit interaction acting on the same
electrons which are responsible for the magnetism. In this
context, a pyrochlore molybdate system $R_2$Mo$_2$O$_7$ ($R=$Nd
and Gd) offers an ideal laboratory to address this
issue~\cite{Taguchi01}.

The both compounds have a unit cell composed of a tetrahedron for
$R$ ions and that for Mo which are shifted from each other by half
the lattice constant. Nd$_2$Mo$_2$O$_7$ henceforth referred to as
the NMO is an itinerant ferromagnet with Curie temperature
$T_C\sim90$ K. By a small amount of hole doping, the marginally
insulating Gd$_2$Mo$_2$O$_7$ turns to be a ferromagnetic metal
with $T_c\sim 60$ K in the case of
(Gd$_{0.95}$Ca$_{0.05}$)$_2$Mo$_2$O$_7$ (henceforth referred to as
the GMO). The ferromangetism in these compounds is mostly
attributed to the double exchange mechanism for mobile carriers of
Mo $4d$ electrons. Well below $T_c$ ($\sim40$ K for $R$=Nd),
through the exchange coupling $J_{fd}$ with the Mo spins, the
localized $4f$ moments on the rare-earth ions $R$=Nd and Gd start
to produce a net magnetic moment in the anti-parallel and the
parallel directions to the ferromagnetic moments of Mo spins,
respectively. Here, it is important to notice that the Nd moments
are subject to a strong single-ion anisotropy and the four moments
in the Nd tetrahedron behave as Ising spins pointing toward or
outward the center of the tetrahedron as in spin-ice
systems~\cite{Taguchi01}. In contrast, the Gd moments behave as
Heisenberg spins since the orbital moment is quenced in the
Gd$^{3+}$ ion with the $4f^7$ configuration~\cite{Taguchi03}. The
noncoplanar spin anisotropy and structure of the Nd moments are
transmitted to the Mo spins through the exchange coupling
$J_{fd}$, leading to the spin chirality of the Mo electrons only
in the case of the NMO.

This spin chirality manifests itself as a gauge flux which is
experienced by the electrons upon their coherent and incoherent
propagations, or the optical $d \to d$ transition between the Mo
atoms, while the local charge transfer excitations on the Mo$-$O
bonds are least influenced by the spin chirality since the oxygen
ion has no spin moment. In the dc limit, the spin chirality
actually contributes to the AHE. In particular, in the case of the
NMO, we can distinguish a \emph{spin chirality induced term} in
$\sigma_{\rm Hall}$ from the conventional one associated with the
spin-orbit coupling of the itinerant electrons. This is because
the spin chirality of the Mo conduction electrons appears without
the spin-orbit interaction acting on themselves. Similarly, it is
expected that the \emph{spin chirality} gives an additional
contribution to $\sigma_{xy}(\omega)$ for the NMO, as far as the
optical $d \to d$ transitions between the Mo atoms is concerned.
Namely, the dc and the optical Hall conductivities of the NMO in
the energy range from $\omega=0$ to the optical $d \to d$
transitions contain the \emph{spin chirality induced
contributions}. On the other hand, those of the GMO, which is a
reference system reproducing well the magnetic and
longitudinal-transport properties of the NMO except the chiral
spin order arise solely from the conventional spin-orbit coupling
acting on the Mo $4d$ electrons. Therefore, the comparison between
NMO and GMO offers a touch-stone for the spin-chirality mechanism.
On this basis, we study magneto-optics due to the spin chirality
from both the theoretical and the experimental viewpoints.

\begin{figure}[h]
\includegraphics[width=8.4cm]{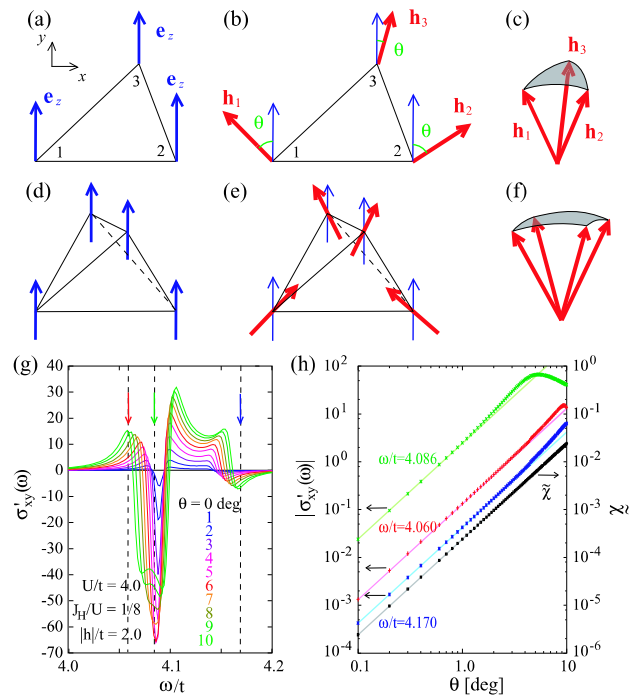}
\caption{(Color) (a) Collinear and (b) noncoplanar magnetic
structures on a triangle cluster with the polar angle $\theta$,
and (c) the associated solid angle. (d) Collinear and (e)
noncoplanar magnetic structures on a tetrahedron, and (f) the
associated solid angle. (g) $\sigma'_{xy}(\omega)$ as a function
of $\theta$ for the two-band Hubbard model on a triangle cluster.
(h) Scaling plot of $|\sigma'_{xy}|$ vs $\theta$ at three energies
denoted by the arrows in (g). Lines are fits to the $\theta^2$
dependence.} \label{fig:theory}
\end{figure}

The Mo$^{4+}$ ion has two $4d$ electrons in the $t_{2g}$ orbital
in the local coordinate associated with each MoO$_6$ octahedron.
The trigonal crystal field splits the triply degenerate $t_{2g}$
orbitals into a single $a_{1g}$ and doubly degenerate $e_g$ orbitals.
One electron occupied in the $a_{1g}$ orbital is more localized
than the other in the $e_g$  orbital~\cite{Solovyev03,Kezsmarki04}.
To mimic this situation of NMO and GMO and to gain fundamental
magneto-optical properties induced by the spin chirality,
we consider the following simplest model Hamiltonian on a triangle cluster;
\begin{eqnarray}
  &&{\cal H}=\sum_{i=1}^3 \big[-t\sum_{m,\alpha}(c^\dagger_{i,m,\alpha}c_{i+1,m,\alpha}+{\rm h.c.})-{\bf h}_i\cdot\sum_m{\bf s}_{i,m}
    \nonumber\\
  &&+U\sum_mn_{i,m,\uparrow}n_{i,m,\downarrow}+U'n_{i,1}n_{i,2}
    -2J_{\rm H}{\bf s}_{i,1}\cdot{\bf s}_{i,2}\big].
\label{eq:H}
\end{eqnarray}
Here, $c_{i,m,\alpha}$ and $c^\dagger_{i,m,\alpha}$ are the
annihilation and creation operators of electron at the site $i$
with the orbital $m$ ($=1$ or $2$) and the spin $\alpha$
($\uparrow$ or $\downarrow$), where the site 4 is identified with
1. (See Figs. 1~(a) and (b).) We have also introduced
$n_{i,m,\alpha}\equiv c^\dagger_{i,m,\alpha}c_{i,m,\alpha}$,
$n_{i,m}\equiv \sum_{\alpha}n_{i,m,\alpha}$, and ${\bf
s}_{i,m}\equiv
\frac{1}{2}\sum_{\alpha,\alpha'}c^\dagger_{i,m,\alpha} {\bf
\sigma}_{\alpha\alpha'}c_{i,m,\alpha'}$. $t$ is an
orbital-diagonal transfer integral between the nearest neighbors,
$U$ and $U'=U-(5/2)J_{\rm H}$ are the local Coulomb interaction
parameters, and $J_{\rm H}$ is the Hund coupling henceforth
assumed to be $U/8$. The second term in the square bracket in
Eq.~(\ref{eq:H}) mimics the molecular fields created by the
$a_{1g}$ and the surrounding $e_g$ electrons as well as the
surrounding rare-earth moments in the two-in two-out configuration
in the context of the spin ice~\cite{Bramwell01}.
Figure~\ref{fig:theory} displays the local spin configuration on a
Mo tetrahedron which is the magnetic unit cell of the Mo
sublattice in pyrochlore-type molybdates. The collinear case in
Fig.~\ref{fig:theory}~(d) corresponds to GMO while the chiral spin
order, the so-called ``umbrella'' structure realized in NMO is
shown in Figs.~\ref{fig:theory}~(e) and (f).

With the spin configurations as shown in Figs.~\ref{fig:theory}~(a)
and (b), we have numerically calculated the conductivity tensor
$\sigma_{\mu\nu}(\omega)$ of this model.
When ${\bf h}_i$'s are collinear as in Fig.~\ref{fig:theory}~(a)
or even coplanar within the $xy$ plane and hence so is the spin alignment,
we obtain $\sigma_{xy}(\omega)=0$.
Now, we introduce the noncoplanar umbrella-like configuration
for ${\bf h}_i$ as depicted in Fig.~\ref{fig:theory}~(b),
where ${\bf h}_i=h{\bf e}_i$ with
${\bf e}_1=(-\frac{\sqrt{3}}{2}\sin\theta,-\frac{1}{2}\sin\theta,\cos\theta)$,
${\bf e}_2=( \frac{\sqrt{3}}{2}\sin\theta,-\frac{1}{2}\sin\theta,\cos\theta)$
and ${\bf e}_3=(0,\sin\theta,\cos\theta)$ in the range of
$0<\theta<90^\circ$. In this case, the spin chirality
$\chi=\langle{\bf S}_1\times{\bf S}_2\cdot{\bf S}_3\rangle$ emerges,
where ${\bf S}_i={\bf s}_{i,1}+{\bf s}_{i,2}$ is the total spin on the site $i$.

Figure~\ref{fig:theory}~(g) represents the main structure in the
real part $\sigma'_{xy}(\omega)$ of the optical Hall conductivity
$\sigma_{xy}(\omega)$ for $0\le\theta\le10^\circ$. For the
collinear case $\theta=0$ and the coplanar case $\theta=90^\circ$
(not shown), $\sigma_{xy}(\omega)$ vanishes. With increasing
$\theta$ from $0$, the $\sigma_{xy}(\omega)$ gradually emerges.
For the present study of the finite-size cluster which is
inherently insulating, the energy scale of $\omega/t\approx4$ is
focused. This spectral region corresponds to incoherent
excitations in the case of the bulk system. The $\theta$
dependence of the spectral intensity at three different energies
are shown in Fig.~\ref{fig:theory}~(h) together with the
normalized spin chirality $\tilde{\chi}=\chi/(\langle
S_1\rangle\langle S_2\rangle\langle S_3\rangle)$. Here, it is
evident that both $\sigma'_{xy}(\omega)$ and $\tilde{\chi}$ are in
proportion to $\theta^2$, and hence
$\sigma_{xy}(\omega)\propto\tilde{\chi}$ for such small values of
$\theta$. As we increase $U$, the optical Hall conductivity
spectra shift to higher energy with the intensity being decreased.
It is notable that a small spin canting affects the
magneto-optical response at such incoherent regime. The on-site
Coulomb interactions only determine the characteristic excitation
energy $\omega$ for the incoherent optical process and the
longitudinal current response at this energy with the factor
$\sim1/\omega$. The ratio of the anomalous Hall velocity to the
longitudinal one is exclusively determined by the tilting angle
$\theta$ and insensitive to the Coulomb interaction parameters.
The triangular-cluster model of non-coplanar spins is the most
basic system to investigate the effect of spin chirality on the
optical Hall conductivity. Although these results allow at most
semi-quantitative comparison with the experiments they generally
prove the importance of the spin-chirality contribution to
$\sigma_{xy}(\omega)$ in the incoherent range of the spectrum (for
intraband transitions or excitaions through the gap) independently
of the fine details of the model system.

Now we turn to the experimental results. Figure~2 shows the
diagonal conductivity spectrum of NMO in the ferromagnetic
metallic ground state. As evidenced by a systematic study of the
bandwidth controlled metal-insulator transition in pyrochlore
molybdates \cite{Kezsmarki04}, the $\omega\leq2$\,eV excitations
correspond to the Mo $4d$ intraband optical processes. The
low-energy side of the conductivity spectrum is well fitted by a
Drude term (with a scattering rate of $\gamma\approx 35$\,meV)
when the contribution of the infrared active phonon modes is
subtracted. The coherence peak is followed by the so-called
incoherent band (or intraband) transitions in the mid-infrared
region which is characteristic of correlated electron systems in
the vicinity of a metal to insulator transition. Separated from
the Mo $4d$ optical processes, the higher energy conductivity peak
(centered around $\sim4.4$\,eV) is assigned as local
charge-transfer excitations from oxygen sites to neighboring Mo
sites.

The dominance of the spin chirality mechanism over the spin-orbit
interaction in the dc AHE of pyrochlore molybdates (namely $\sim
2$ orders of magnitude difference) has been previously pointed out
in frame of a comparative investigation of NMO and GMO
\cite{Taguchi01}. Now, on the basis of the present wide
energy-range magneto-optical experiments we shall conclude that
the same tendency remains valid for the finite-frequency
transverse transport, $\sigma_{xy}(\omega)$, as far as the Mo $4d$
intraband-transition region is concerned.

Figure~3 gives a comparison between the optical properties of NMO
and GMO in their ground state. The measured quantities are shown
on the left panel, while real and imaginary parts
($\sigma'_{\mu\nu}(\omega)$ and $\sigma''_{\mu\nu}(\omega)$,
respectively) of the conductivity tensor $\sigma_{\mu\nu}(\omega)$
on the right. They are plotted on a common energy scale over the
range of the magneto-optical study. $\sigma'_{xx}(\omega)$ was
obtained by Kramers-Kronig transformation from the reflectivity,
while due to the applied ellipsometric method,
$\sigma_{xy}(\omega)$ could be directly determined from the
magneto-optical Kerr parameters:
\begin{equation}
\sigma_{xy}(\omega)=-(\Theta_K(\omega)+i\eta_K(\omega))\sigma_{xx}(\omega)\sqrt{1+4\pi
i\sigma_{xx}(\omega)/\omega}\ ,
\end{equation}
where $\Theta_K(\omega)$ and $\eta_K(\omega)$ are the Kerr rotation
and the Kerr ellipticity, respectively, measured by a polarization
modulation technique~\cite{Ogasawara}.
For both compounds, with nearly normal incidence,
the polar Kerr spectra of the $(1 0 0)$ crystallographic plane
were measured in a magnetic field of $B=0.26$\,T
in the temperature range of $T=10-120$\,K. This magnetic field is
large enough to orientate the ferromagnetic domains
but still can be considered as the zero-field limit.

Both compounds can be classified as bad metals since even at the
lowest temperature the resistivity is close to or even larger than
the Ioffe-Regel limit ($\rho_{_{I-R}}\approx0.53$\,m$\Omega$cm)
(see Fig.~3). The whole $\sigma_{xx}$ spectra are also similar to
each other in magnitude. By contrast, we have observed large
signals $\Theta_K$ and $\eta_K$ for NMO below $\sim2$\,eV, while
substantially smaller ones for GMO. In both cases, there appear a
peak structure in $\Theta_K(\omega)$ and the corresponding
inflection in $\eta_K(\omega)$ centered around $\sim1$\,eV, but
these are due to the presence of the apparent plasma edge
(indicated by a dashed line). After evaluating
$\sigma_{xy}(\omega)$ according to Eq.~(2) the apparent plasma
resonance is naturally cancelled out by the energy dependence of
$\sigma_{xx}(\omega)$, which ensures the correctness of the
Kramers-Kronig transformation. Nevertheless, a significant
difference is quite clear in the magnitude of
$\sigma_{xy}(\omega)$ for the infrared incoherent band
($\omega\lesssim1$\,eV) between the two compounds.

On the high-energy side which is dominated by the O $2p$
$\rightarrow$ Mo $4d$ charge-transfer excitations, the
magneto-optical response of the two systems are almost identical.
This suggests that in this energy region, $\sigma_{xy}(\omega)$
solely originates from the spin-orbit interaction of the Mo
electrons and simply measure the spin polarization of the final
state (not the chirality), therefore invariant under the change of
the rare-earth component. Towards lower energies where the Mo $4d$
intraband transitions are dominant, however, $\sigma_{xy}(\omega)$
shows a robust increase in the case of NMO while not for GMO. At
the lowest energy of the experiment, $\omega=0.11$\,eV, this
tendency results in nearly one order of magnitude difference
between the magneto-optical responses of the two compounds, as
expressed by $\sigma'_{xy}({\rm NMO})/\sigma'_{xy}({\rm
GMO})\approx6.5$ and $\sigma''_{xy}({\rm NMO})/\sigma''_{xy}({\rm
GMO})\approx11.4$.

In Fig.~4, we present the detailed temperature dependence of
$\sigma_{xy}(\omega)$ for NMO. The spectral structure does not
vary with increasing temperature, but only its magnitude is
suppressed as $T_c$ is approached. Figure~4 shows the temperature
dependence of $\sigma_{xy}(\omega)$ for the both compounds at
three representative energies: (i) in the dc limit, (ii) in the
region of the incoherent intraband transitions ($\omega=0.11$\,eV)
and (iii) for the charge-transfer peak ($\omega=4.4$\,eV). The
total magnetization and the Mo-spin contribution~\cite{Taguchi01}
are also plotted in the same figure. Below $50$\,K, the Nd moments
tend to align antiparallel to the almost ferromagnetically ordered
Mo spins which results in the low-temperature reduction of the
magnetization. By contrast, the interaction between Mo and Gd
sites is ferromagnetic, and hence the magnetization is enhanced in
GMO. As far as the Mo $4d$ intraband (incoherent band) transitions
at $0.1-1$\,eV are concerned, $\sigma_{xy}(\omega)$ of NMO is
substantially larger than that of GMO over the whole ferromagnetic
phase of the compounds, independently of temperature. This key
finding evidences that the leading term in the magneto-optical
response of the incoherently moving electrons comes from the
spin-chirality mechanism.

Generally speaking, the role of the spin chirality becomes more
important as the dc limit is approached, since the low-energy
transitions can more coherently pick up the gauge flux (or
integrate the Berry curvature). When the Fermi level is located
within a tiny energy gap generated by the gauge flux at the band
crossing point, $\sigma_{xy}(\omega$$=$$0)$ is maximized due to
the presence of the magnetic-monopole analog at the crossing point
\cite{Fang03}. In the low-temperature phase of NMO, this
characteristic energy scale is estimated to be $E_0 \sim 0.02$\,eV
\cite{note}. The temperature dependence of $\sigma_{xy}(\omega)$
should show a complicated behavior for $\omega < E_0$ as a finger
print of this gauge flux structure in momentum space. For $\omega
\gg E_0$, on the other hand, the perturbative treatment of the
magnetization $M$ times the spin chirality $\chi$ is appropriate,
and hence we predict $\sigma_{xy} \propto \theta^2 M$. Therefore,
considering $E_0< 0.1$\,eV, it is reasonable that the temperature
dependence of $\sigma_{xy}(\omega)$ in Figs.~4 and 5 scales with
the Mo magnetization except the dc value.

A nonmonotonous spectral shape of $\sigma_{xy}(\omega)$ is
expected for $\omega<0.1$\,eV, since it is known that complicated
band crossings realized in the electronic structure gives a rapid
oscillation of $\sigma_{xy}(\omega)$ as mentioned above. Indeed,
it is discerned in the low-energy $\sigma_{xy}(\omega)$ spectra of
NMO. Furthermore, $\sigma''_{xy}(\omega)$ obviously changes sign
at $\sim0.1 eV$. From the Kramers-Kronig relation, this has to be
accompanied by a maximum in $\sigma'_{xy}(\omega)$. Similarly, one
can show that the compatibility of the dc values
($\sigma'_{xy}(\omega=0) > 0$ and $\sigma''_{xy}(\omega) = 0$)
with the low-energy $\sigma_{xy}(\omega)$ data presented in
Fig.~4, $\sigma'_{xy}(\omega)$ requires two more sign changes in
the experimentally inaccessible low-energy region as the dc limit
is approached. Such a behavior is consistent with the above
scenario of the rapid oscillation of $\sigma'_{xy}(\omega)$ at low
energies due to the band crossing.

In conclusion, we have demonstrated, theoretically and
experimentally, the magneto-optical effect induced by the
spin chirality. The theory on the two-band Hubbard model for a
triangle cluster shows that the optical Hall conductivity across
the Mott-Hubbard gap emerges in the presence of the spin chirality.
The magneto-optical experiments have shown that the amplitude of the
optical Hall conductivity in the energy range of $d \to d$ intraband
transitions between the Mo atoms is large under a noncoplanar spin structure
in Nd$_2$Mo$_2$O$_7$ with spin chirality, while it is significantly
suppressed in Gd$_2$Mo$_2$O$_7$ without spin chirality.

This work was in part supported by a Grant-In-Aid for Scientific
Research from the MEXT, Japan. I.\ K.\ acknowledges support from
JSPS.

%
%

%
%
\section*{Figure Captions}
%
%
\begin{figure}[h]
\includegraphics[width=3.3in]{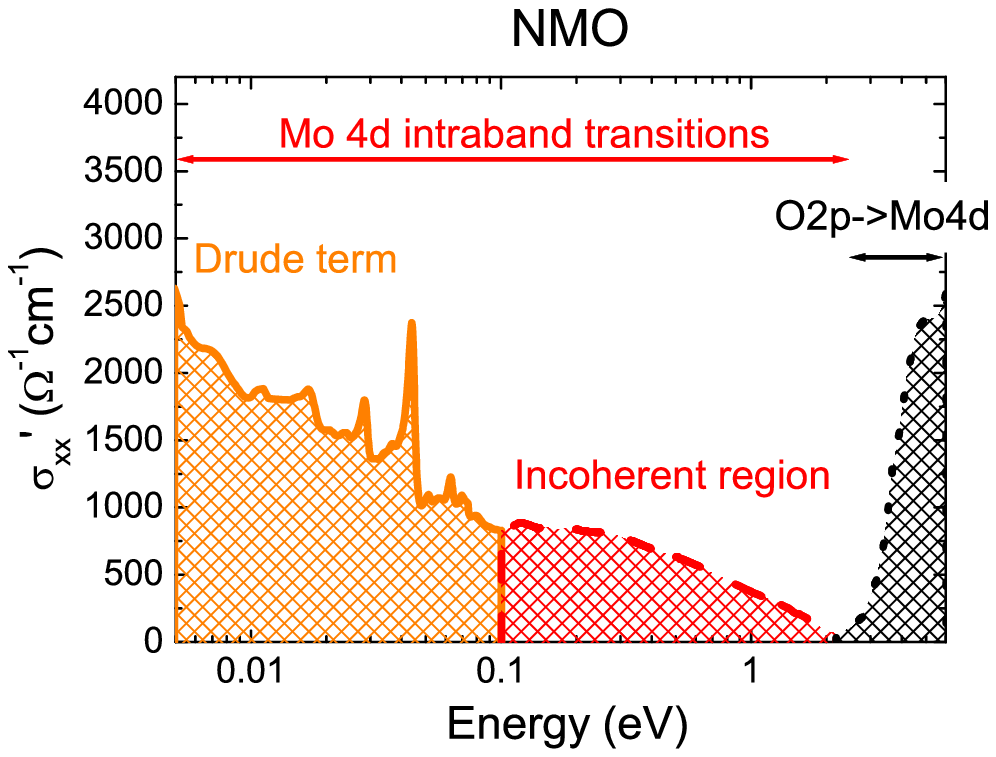}
\caption{(Color online) Real part of the diagonal conductivity
$\sigma'_{xx}(\omega)$ in the ground state of NMO. The region of
the Mo $4d$ intraband transitions and the O $2p$ $\rightarrow$ Mo
$4d$ charge-transfer excitations are indicated. The former is
further divided into two parts: Drude term and mid-infrared
incoherent excitations.} \label{fig:mo}
\end{figure}
%
\begin{figure}[h]
\includegraphics[width=3.3in]{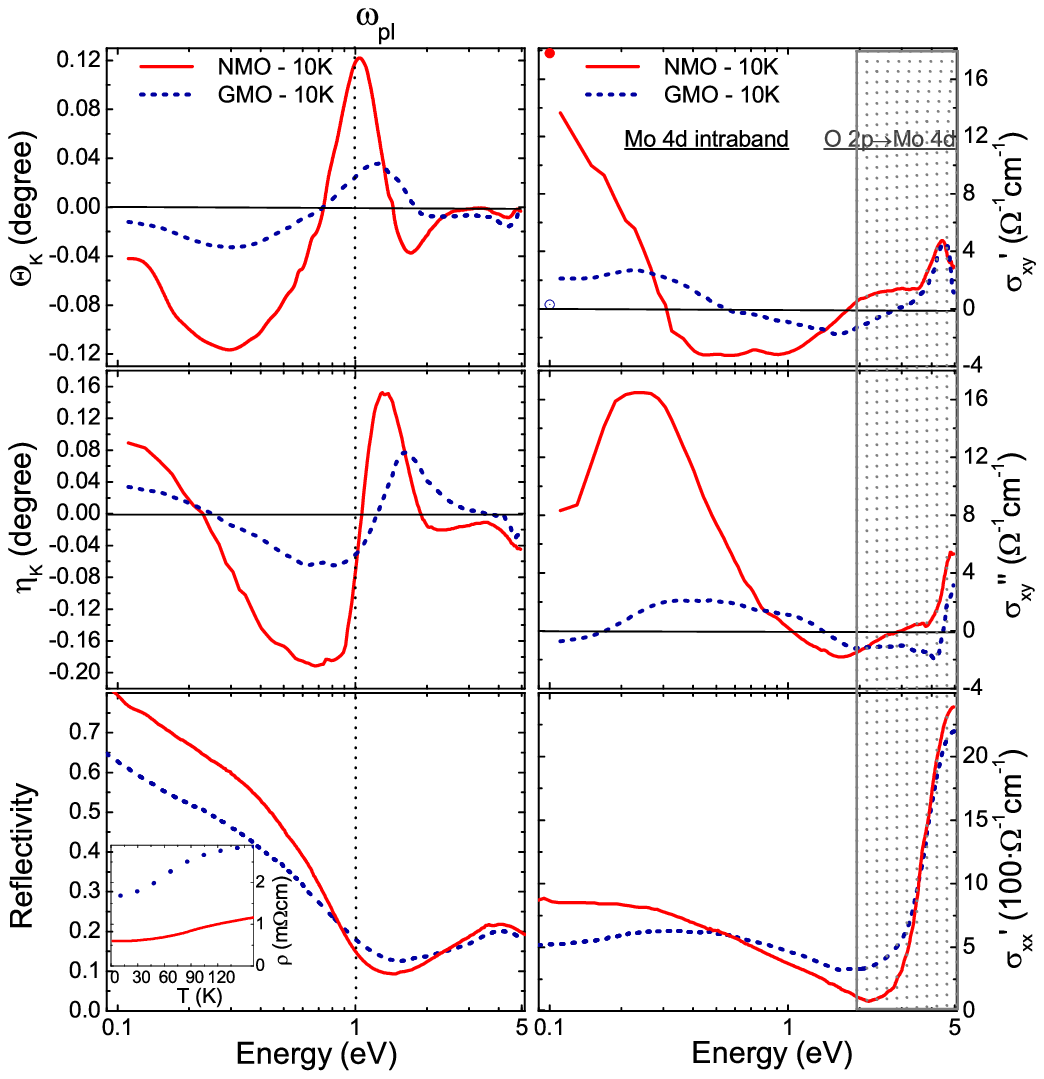}
\caption{(Color online) Left panels: Kerr rotation $\Theta_K$,
Kerr ellipticity $\eta_K$ and reflectivity spectra of
Nd$_2$Mo$_2$O$_7$ (NMO) and
$($Gd$_{0.95}$Ca$_{0.05}$$)$$_2$Mo$_2$O$_7$ (GMO) measured at
$T=10$\,K. The dotted line running through the plots indicates the
plasma edge. The inset in the bottom panel shows the comparison of
the dc resistivity for the two materials below $T=160$\,K. Right
panels: Spectra of $\sigma'_{xy}$, $\sigma''_{xy}$, and
$\sigma'_{xx}$ at 10\,K. In the top panel, the dc values
$\sigma_{\rm Hall}$ are also indicated by dots. The high-energy
region dominated by charge-transfer excitations is distinguished
from the Mo $4d$ intraband transitions by a gray pattern.}
\label{fig:mo}
\end{figure}
%
\begin{figure}[h]
\includegraphics[width=2.3in]{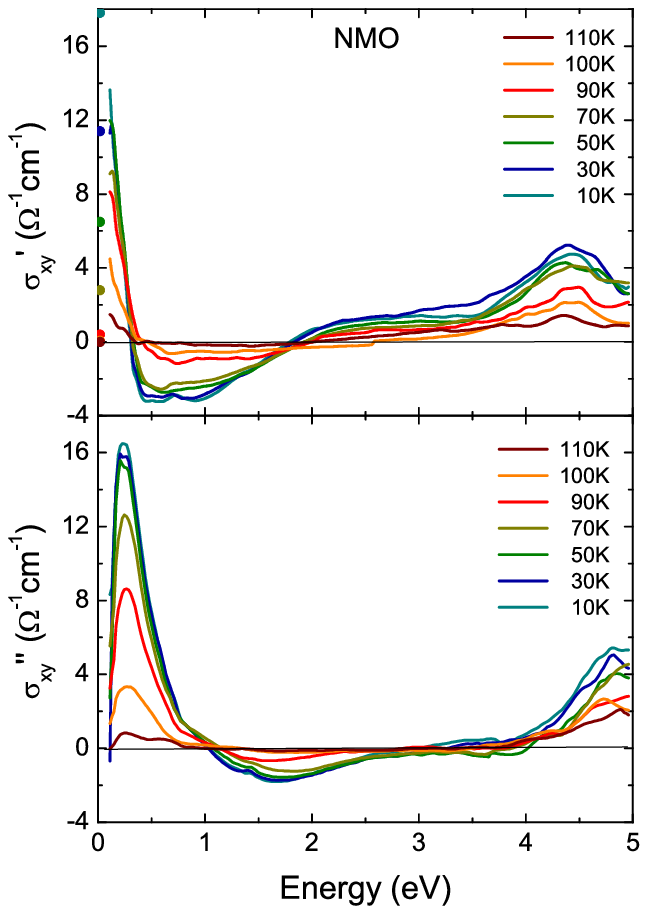}
\caption{(Color) Spectra of $\sigma'_{xy}$ and $\sigma''_{xy}$ of
NMO at various temperatures. The corresponding dc values
($\sigma_{\rm Hall}$) are also plotted with dots.}
\label{fig:sigmaxy_E}
\end{figure}
%
\begin{figure}[h]
\includegraphics[width=2.6in]{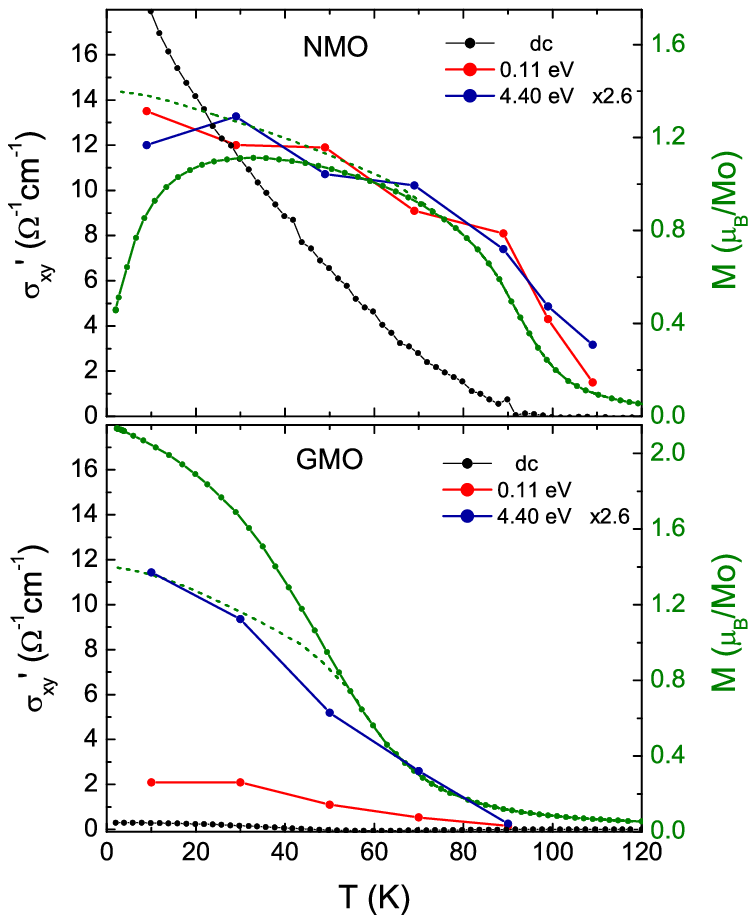}
\caption{(Color) The temperature dependences of
$\sigma_{xy}'(\omega)$ and the magnetization both for NMO and GMO.
$\sigma_{xy}'(\omega,T)$ is shown for three characteristic
energies, $\omega=0$, $0.11$ and $4.4$\,eV by black, red and blue
symbols, respectively. Note that the signal at $4.4$\, eV is
multiplied with a factor of $2.6$. The overall magnetization is
indicated by green symbols while the contribution of the Mo spins
alone by a dashed line.} \label{fig:sigmaxy_T}
\end{figure}

\begin{references}
\bibitem{note}
 In the simplest model $E_0$ can be expressed as bandwidth times
$(\theta/\pi)^2$ ($\theta$: tilting angle of the spins). According
to the estimation $\theta \cong 5^\circ$ \cite{Taguchi01,Yasui01},
we have $E_0 \sim 0.002$\,eV. However, the unusual temperature
dependence of $\sigma_{xy}(\omega$$=$$0)$, observed all the way up
to $T_c$, implies that $E_0$ is about one order of magnitude
larger.
\end{references}
\end{document}